\documentclass{pasj00}
\begin{document}
\SetRunningHead{Kirihara, Miki \& Mori}{Outer Density Profile of M31's Dark Matter Halo}
\Received{xxxx/xx/xx}
\Accepted{yyyy/yy/yy}

\title{
Puzzling Outer--Density Profile of the Dark Matter Halo in the Andromeda Galaxy
}

\author{
Takanobu \textsc{Kirihara}\altaffilmark{},
Yohei \textsc{Miki}\altaffilmark{}
and
Masao \textsc{Mori}\altaffilmark{}}
\altaffiltext{}{University of Tsukuba, Tennodai 1-1-1, Tsukuba, Ibaraki, Japan}
\email{kirihara@ccs.tsukuba.ac.jp
}

%

\maketitle

%
%
%
%

\begin{abstract}

The cold dark matter (CDM) cosmology, which is the standard theory of the structure formation in the universe, predicts that the outer density profile of dark matter halos decreases with the cube of distance from the center. 
However, so far not much effort has examined this hypothesis. 
In the halo of the Andromeda galaxy (M31), large--scale stellar structures detected by the recent observations provide a potentially suitable window to investigate the mass--density distribution of the dark matter halo. 
We explore the density structure of the dark matter halo in M31 using an $N$--body simulation of the interaction between an accreting satellite galaxy and M31. 
To reproduce the Andromeda Giant Southern Stream and the stellar shells at the east and west sides of M31, we find the sufficient condition for the power--law index $\alpha$ of the outer density distribution of the dark matter halo. 
The best--fit parameter is $\alpha=-3.7$, which is steeper than the CDM prediction.
\end{abstract}

%
%
%
%

\section{Introduction}
\label{sec:Introduction}

The cold dark matter (CDM) model as the standard paradigm for structure formation in the universe predicts that galaxies have grown larger through many mergers with less massive galaxies. 
Cosmological simulations of structure formation suggest that the spherically--averaged density profiles of dark matter halos (hereafter DMHs) have a universal shape (\cite{Nav1996apj, Nav1997apj, Die2004mnras} and references therein). 
In these profiles, much controversy exists over the mass--density distribution of the inner DMH and this issue is still not concluded (\cite{1994Natur.370..629M, 1995ApJ...447L..25B, Nav1996apj, Fuk1997apj, Moo1998apj, Nav2004mnras, Ogi2011apj}, 2013; \cite{Ish2013apj} and references therein). 
With respect to the density distribution of the outermost region, however, the mass--density decreases with the cube of the distance from the center of the DMH in most earlier studies. 

In recent years, deviations from the standard power--law index in the outer density profile of DMHs have been discussed using cosmological $N$--body simulations \citep{Nav2010mnras}. 
In observational scope, weak lensing is beginning to probe the outskirts of galaxy clusters \citep{Bri2013mnras}. 
That is to say the outward region of the DMH is the excellent laboratory to examine the prediction of the CDM model. 
However, because the stellar and/or gas density is too low to detect even with the latest instruments, it is extremely difficult to measure the mass-density distribution of the outer region of the galaxy. 
Consequently, the theoretical prediction has not yet been carefully tested by observations. 

Only recently, deep and panoramic surveys of the Andromeda galaxy (M31), which is the nearest large galaxy located $780$ kpc away from our Galaxy \citep{Fon2006aj}, revealed that the outer region of M31 shows a wealth of stellar substructures (\cite{Gil2009apj,Mar2013apj}). 
The Andromeda Giant Southern Stream (hereafter GSS) extends about $120$ kpc further away along the line of sight from M31 \citep{Iba2001nat,McC2003mnras}, and the radial--velocity distribution has been observed (\cite{Iba2004mnras,Gil2009apj}). 
The large--scale stellar shells at the east and west sides of M31 spread like a fan with a constant radius about $30$ kpc from the center of M31 (\cite{Gil2009apj,Mar2013apj}). 
These structures have been thought the evidence of a galaxy collision \citep{Far2007mnras,Mor2008apj,Ham2010apj}. 
Earlier studies examined the orbits and the mass of an accreting satellite galaxy, which reproduce all of these structures \citep{Fon2006aj,Far2006mnras,Far2007mnras,Mor2008apj}. 

These structures spread far beyond the stellar disk in M31 and are attractive for the investigation of the actual density distribution of the DMH in M31. 
Nevertheless, earlier studies have mostly assumed that the density profile of the outer DMH of M31 decreased with the cube of the distance from the center $\rho(r)\propto r^{-3}$ in accordance with the CDM prediction. 
No study has yet examined the density distribution of the outer DMH in M31. 
These situations motivate us to test the CDM prediction of the density profile of the DMH, using the $N$--body simulation for the formation of the GSS and the stellar shells in the DMH with the different density profiles. 
We describe our numerical model in \S 2 and present simulation results in \$ 3. A brief summary and discussion are stated in \S 4.

%
%
%
%

\section{Numerical model}

To explore the density profile of the DMH in M31,  we demonstrate the interaction between an accreting satellite dwarf galaxy and M31 using the $N$--body simulations. 
So far, the self--gravitating response of the bulge, disk, and DMH of M31 to an accreting satellite galaxy was studied by \citet{Mor2008apj}, and they concluded that satellite galaxies less massive than $5\times10^9\MO$ had a negligible effect on the gravitational potential of M31. 
Accordingly, we simply assume M31 as the source of a fixed gravitational potential composed of a bulge, a disk, and a DMH. 

The disk of M31 is adopted an exponential disk with the radial scale length of $5.4$ kpc, the scale height of $0.60$ kpc and the mass of $3.66\times 10^{10}$\MO. 
The bulge is assumed to have a spherically symmetric distribution represented by a Hernquist profile \citep{Her1990apj} with the scale radius of $0.61$ kpc and the mass of $3.24\times 10^{10}$\MO. 
This model nicely reproduces the profile of surface brightness of the M31 disk and bulge, and the velocity dispersion of the bulge \citep{Gee2006mnras,Far2007mnras}. 

The Navarro--Frenk--White (NFW) model that is widely accepted density profile of CDM halos is empirically derived from cosmological $N$--body simulations \citep{Nav1996apj,Nav1997apj}. 
The resultant profiles of density distribution are approximately fitted by 
 $\rho_{\mathrm{NFW}}(r)=\rho_{s}(r/r_{s})^{-1}(1+r/r_{s})^{-2}$, 
 where $r_{s}$ and $\rho_{s}$ are the scale radius and the scale density, respectively. 
\citet{Far2007mnras} adopted $r_s=7.63$ kpc and $\rho_s=6.17\times 10^7 \MO ~{\mathrm {kpc}}^{-3}$. 
We focus on making a diagnosis of the density profile in the outer CDM halo using $N$--body experiments to reproduce the GSS. 
In this purpose, we extend the equation and introduce a power--law index $\alpha(<-2.0)$ in the equation of the density distribution as 
$\rho_{\mathrm{DMH}}(r)=\rho_{s,\; \alpha}(r/r_{s,\; \alpha})^{-1}(1+r/r_{s,\; \alpha})^{\alpha+1}$, 
where $r_{s, \alpha}$ and $\rho_{s, \alpha}$ are also the scale radius and the scale density, respectively. 
In this equation, outer density profile of the DMH approaches asymptotically to the simple power--law distribution $\rho_{\mathrm{DMH}}(r) \propto r^{\alpha}$.

\begin{figure}[htbp]
 \begin{center}
  \FigureFile(80mm,80mm){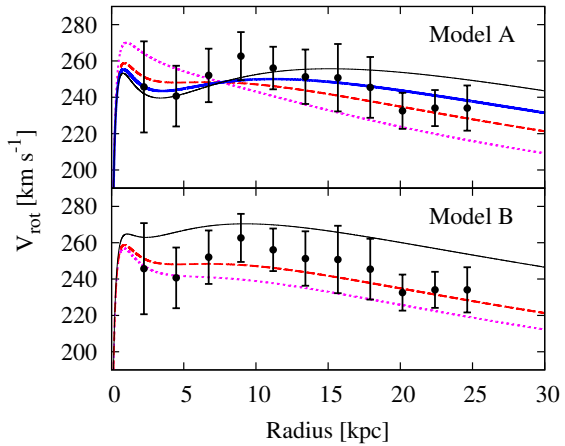}
  \end{center}
  \caption{\label{fig:rotation}
Rotation curves of our M31 model. 
Each line in Model A represents $\alpha=-2.5$ (magenta dotted line), $\alpha=-3.0$ (red dashed line), $\alpha=-3.7$ (blue thick solid line), and $\alpha=-5.5$ (black thin solid line), respectively. 
Each line in Model B represents $\rho_{s, -3}=0.9 \rho_s$ (magenta dotted line), $\rho_{s, -3}= \rho_s$ (red dashed line), and $\rho_{s, -3}=1.3 \rho_s$ (black thin solid line), respectively. 
The observational data are taken from \citet{Ken1989pasp} and \citet{Bra1991apj}.
}
\end{figure}

In this paper, we examine the dependence of power--law index (Model A) and the total mass of the DMH (Model B). 
In Model A, we consider the parameter $\alpha$ ranging from $-6.0$ to $-2.3$ by $0.1$ ($\alpha=-3.0$ corresponds NFW profile). 
It is assumed that the enclosed masses of the DMH at $r_{s}$ and $R=125~\rm{kpc}$ are fixed to $6.66\times 10^{10}\MO$ and $6.59\times 10^{11}\MO$, respectively. 
In Model B, we change the scale density $\rho_{{s,-3}}$ from $0.5\rho_{s}$ to $2\rho_{s}$, keeping the power--law index $\alpha=-3.0$ and the scale radius $r_s$. 
Figure \ref{fig:rotation} shows the comparison of the rotation curves of M31 between the observation and our models. 
It is clear that our models reasonably fit the observation. 

Rotation curve is a key to determine the outer structure of the DMH, but the observations are available only at the inner part of M31 compared to the size of the DMH (a few times of $r_s$). 
Therefore, the GSS is a suitable site for exploring the outer density profile of the DMH of M31. 

To study the dynamical response of the orbiting satellite galaxy, we adopt the satellite as a Plummer's sphere with total mass $2.2\times10^9 \MO$ and scale radius $1.03$ kpc. 
It is generated by a self--consistent $N$--body realization with $245,760$ particles, and the initial position vector and velocity vector for the standard coordinates centered on M31 and detailed satellite model are taken from \citet{Far2007mnras}. 
In Model A, we also perform low--resolution runs for a convergence test with $49,152$ particles, which is $1/5$ times fewer particles of our high--resolution runs.

\begin{figure}[htbp]
  \begin{center}
   \FigureFile(85mm,85mm){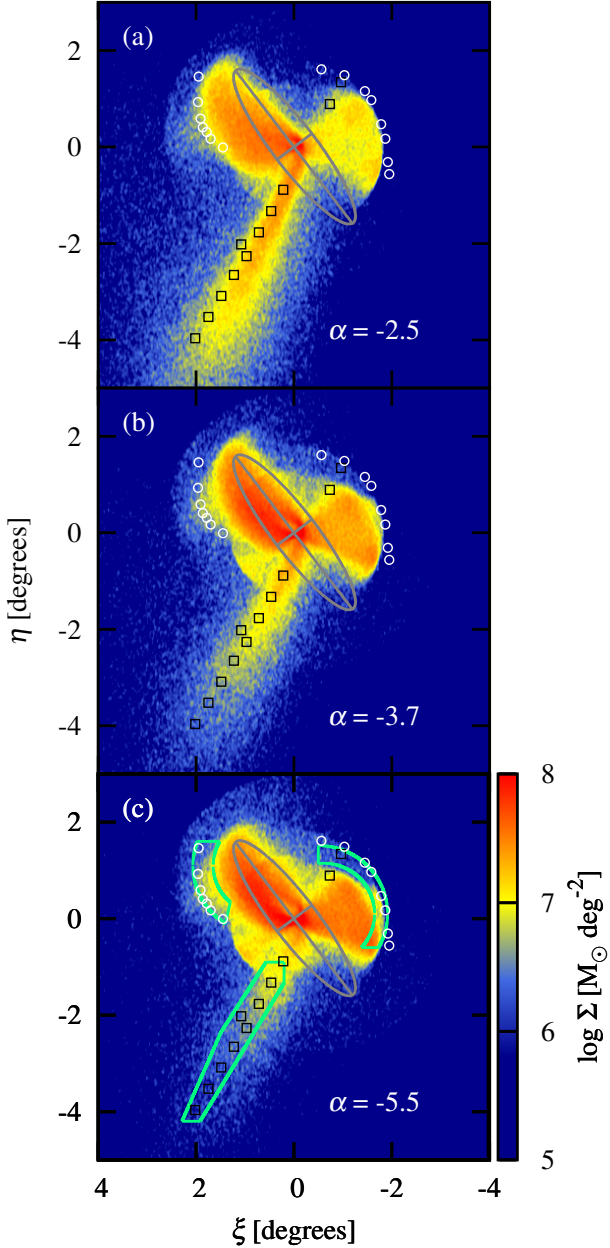}
  \end{center}
  \caption{\label{label1}
Surface mass--density distribution of the satellite galaxy at the best--fit epochs in Model A. 
Each panel corresponds to the case for (a) $\alpha=-2.5$, (b) $\alpha=-3.7$, and (c) $\alpha=-5.5$, respectively. 
The positional coordinates $\xi$ and $\eta$ point the eastern and the northern direction on the sky, respectively. 
The coordinate origin is the center of M31 and $1^{\circ}$ in angle corresponds to $13.6$ kpc. 
In each panel, gray ellipse represents the disk shape of M31. 
Black squares and white circles indicate the observational fields of the GSS taken from Table 1 of \citet{Fon2006aj} and the edges of the shells taken from Table 1 of \citet{Far2007mnras}, respectively. 
The color corresponds to the logarithmic surface density $5.0 \leq \mathrm{log} \Sigma (\MO \mathrm{deg}^{-2}) \leq 8.0$. 
The regions enclosed by green lines in Panel (c) indicate the area for the analysis for the surface mass--density ratio. }
\end{figure}
\begin{figure}[htbp]
\begin{center}
  \FigureFile(85mm,85mm){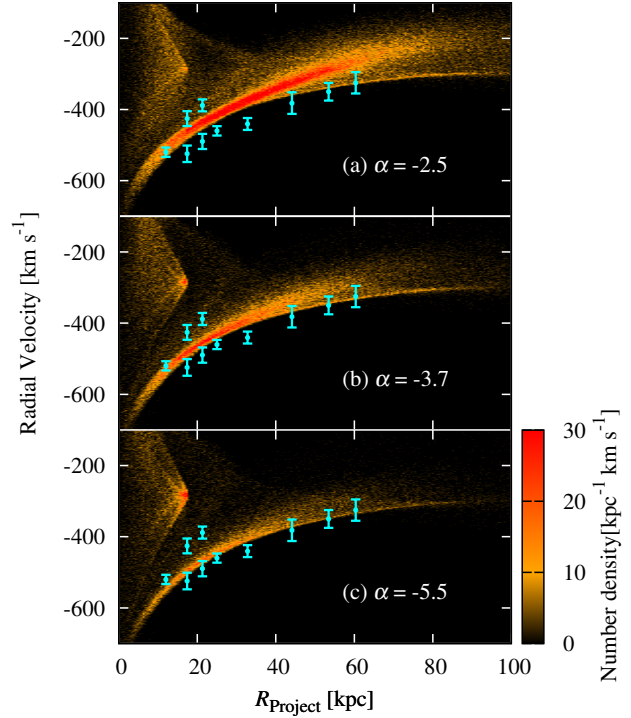}
\end{center}
\caption{\label{label2}Radial velocity distribution of the GSS in Model A. 
The density map shows the results of the $N$--body simulation in the stream region (southeastern area in figure \ref{label1}). 
Blue symbols show the observational data taken from \citet{Fer2004arxiv}, \citet{Kal2006apj}, \citet{Gil2009apj}. 
Each panel corresponds to the case for (a) $\alpha=-2.5$, (b) $\alpha=-3.7$, and (c) $\alpha=-5.5$, respectively. 
}
\end{figure}

\begin{figure*}
 \begin{center}
  \FigureFile(160mm,100mm){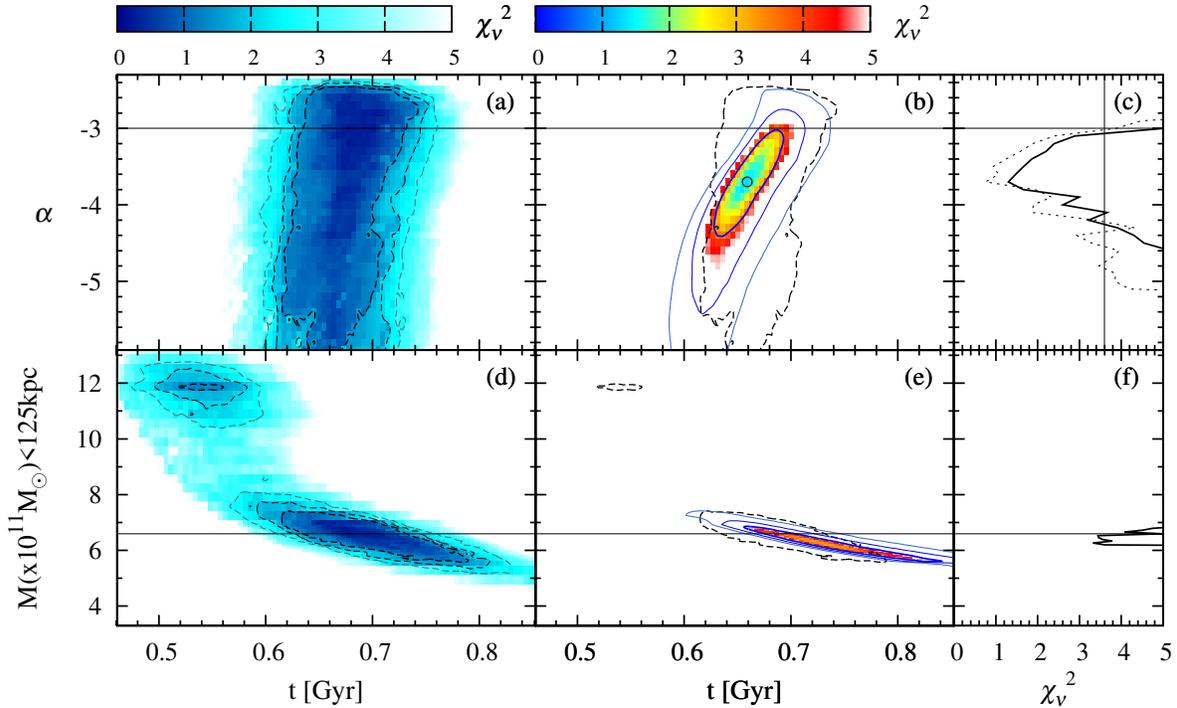}
 \end{center}
\caption{\label{label3}
The results of $\chi^2_{\nu}$ analysis for Model A (upper panels) and Model B (lower panels). 
(a) and (d) are the reduced--$\chi^2$ map of the positions and the shapes of shells. 
(b) and (e) are that of the surface density ratio among the GSS and the two shells. 
(c) and (f) show the minimum $\chi_{\nu}^2$ of the surface density ratio among the GSS and the two shells. 
Each contour describe $1\sigma$ (thick), $2\sigma$ (middle), and $3\sigma$ (thin) confidence intervals of the $\Delta \chi^2_{\nu}$. 
Dashed curves in (b) and (e) correspond to $1\sigma$ confidence level of the positions and the shapes of shells. 
Horizontal line in upper (lower) panels indicates $\alpha=-3$ ($M_{125 \rm{kpc}}=6.59\times 10^{11}M_{\odot}$). 
The solid (dotted) curve in (c) shows the result of high (low)--resolution runs. 
Vertical solid line in (c) corresponds the value of $1\sigma$ confidence interval and the black circle in (b) is the best--fit parameter. 
}
\end{figure*}

%
%
%
%

\section{Simulation}
\label{sec:simlations}

We use an original parallel--tree code with a tolerance parameter of $\theta=0.5$ and a softening length of $60$ pc for all runs. 
Numerical computations have been carried out by T2K--Tsukuba System and HA--PACS in Center for Computational Sciences, University of Tsukuba. 
We compare the observational spatial structures of the stellar stream and two shells and radial velocity distribution of the stellar stream with that of the results of our simulations. 

The result of simulations shows that the satellite collided almost head--on with the bulge of M31, and the first pericentric passage occurred about $0.7$ Gyr ago. 
Then, a large portion of the satellite particles acquires a high velocity relative to the center of M31, and the distribution of satellite particles is spread out significantly. 
This debris expands to a considerable distance keeping a elongated shape and giving rise to the GSS.
Subsequently, particles start to form a clear shell structure at the north--east area of M31. 
Particles move to the west area in M31 and produce the western shell after the formation of the eastern shell. 

Figure \ref{label1} shows the surface mass--density distribution of the debris of the satellite galaxy at the best--fit epochs in Model A. 
It is clear that the smaller $\alpha$ leads to the lower surface density of the GSS. 
This trend mainly comes from the difference of enclosed mass of DMH around the GSS, since the larger enclosed mass corresponds to the shorter free--fall time. 
In the case of a small $\alpha$, stellar particles move quickly to the eastern shell after the formation of the GSS, and then, the GSS has low surface density. 
In other words, the DMH model with a small $\alpha$ further accelerates the dynamical evolution of the debris. 

Figure \ref{label2} shows the radial velocity distribution of the GSS. 
The density map shows the results of $N$--body simulations. 
We assume M31's heliocentric radial velocity of $-300$ km s$^{-1}$ \citep{Fon2006aj}. 
It shows that the radial velocity distributions of the best--fit parameter are consistent with those of observations within the measurement error range. 

For the purpose of quantitative comparison between the observation and the simulation, we compute the reduced--$\chi^2$ of the position of the eastern and the western shell (ES and WS, respectively), and that of the surface density ratio among the GSS and the two shells. 
The reduced $\chi^2$ in each snapshot is given by
\begin{equation}
\chi^2_{\nu}\equiv \frac{1}{\nu}\sum^{N-1}_{i=0}\left(\frac{x_{i, \rm{sim}}-x_{i, \rm{obs}}}{\sigma_{\rm{obs}}}\right)^2,\; \nu=N-1. \label{eq:eq4}
\end{equation}
 We adopt $\sigma_{\rm{obs}}=0.1^{\circ}\sim 1$ kpc which is the maximum edge width of eastern shell estimated from the star count maps obtained by \citet{Irw2005apj}. 
In analyzing the positions of shells, the $1\sigma, 2\sigma$, and $3\sigma$ level corresponds to $\Delta$$\chi_{\nu}^2=1.2, 1.8$, and $2.6$, respectively ($\nu=14$; \cite{Press2007}). 
For analyzing the surface density ratio, we adopt $\Sigma_{\rm{ES}}/\Sigma_{\rm{GSS}}=1.1$ with $\sigma_{\rm obs}=0.10$ and $\Sigma_{\rm{ES}}/\Sigma_{\rm{WS}}=1.1$ with $\sigma_{\rm{obs}}=0.06$ (see also \cite{Irw2005apj}), and the $1\sigma, 2\sigma$, and $3\sigma$ confidence intervals correspond to $\Delta\chi_{\nu}^2=2.3, 6.2$, and $11.8$, respectively ($\nu=2$; \cite{Press2007}). 

Figure \ref{label3} shows the results of $\chi_{\nu}^2$ analysis for Model A and Model B in each time step. 
In Model A, $\chi^2_{\nu}$ of the positions and the shapes of shells is within $1\sigma$ level for $\alpha\lesssim -2.4$ during $0.6-0.7$~Gyr after the start of the simulation runs (see figure \ref{label3}a). 
Similarly, that of the surface density ratio constrains $-4.3<\alpha<-3.0$ (see figures \ref{label3}b and \ref{label3}c). 
Combination with these results, the model of the DMH having the outer density profile with the power--law index of $-4.3 < \alpha < -3.0$ reasonably reproduces these observed structures, and the best--fit parameter is $\alpha=-3.7$. 
This is steeper than the standard CDM prediction ($\alpha=-3.0$). 

Figure \ref{label3}c shows the minimum $\chi_{\nu}^2$ of the surface density ratio among the GSS and the two shells in Model A. 
This figure definitely demonstrates that our result independents on the numerical resolution.

Figures \ref{label3}d and \ref{label3}e show the $\chi^2_{\nu}$ map of the result of the different halo mass model of M31, and figure \ref{label3}f shows the minimum $\chi_{\nu}^2$ of the surface density ratio in Model B. 
The minimum $\chi_{\nu}^2$ value in figure \ref{label3}f ($\chi^2_{\nu}=3.2$) is much larger than that of figure \ref{label3}c ($\chi^2_{\nu}=1.3$). 
It is less likely that the DMH of M31 has the power--law index $\alpha=-3$. 
Therefore, the mass--density distribution of the DMH plays an essential role to reproduce the observational structures than the mass itself.

%
%
%
%

\section{Summary and Discussion}
\label{sec:summary}

We examined the density profile of the DMH of M31 using the $N$--body simulation of the galaxy collision. 
The best--fit parameter of the outer density slope of the DMH is $\alpha=-3.7$ to reproduce the GSS and the shell structures observed in M31. 
This result advocates that the mass--density profile of DMH in M31 is steeper than that of the prediction of the CDM model ($\alpha=-3$). 

In Model A, the simulation for varying $\alpha$ assumed the fixed enclosed mass of DMHs at a radius $R=125$ kpc, which approximately corresponds to the length of the visible GSS. 
We also examined the enclosed mass at $R=195$ kpc, which is the virial radius of M31, and the enclosed mass of the DMH is $7.98\times 10^{11} \MO$ (see also \cite{Gee2006mnras}; \cite{Far2006mnras}). 
Again, we got the similar results, the most suitable parameter is $\alpha=-3.3^{+0.3}_{-0.5}$. 

This discrepancy may arise from several effects which we do not take into account in our model such as the triaxiality of the DMH, the tidal effect of nearby galaxy, the morphology of the accreting satellite, the infalling orbit of the progenitor. 
Actually, cosmological $N$--body simulations suggest that CDM halos are generally triaxial \citep{Jin2002apj,Hay2007mnras}. 
In the case of the Milky Way galaxy, \citet{Deg2013mnras} proposed the triaxiality of DMH interpreting the kinematic data of the Sagittarius stream. 
The tidal effects of M33 and the Milky Way galaxy might change the structure of the DMH in M31. 
M33 is the most massive satellite galaxy of M31 and its virial mass is about $1.5\times10^{11}\MO$ \citep{Mar2012apj}. 
It locates on the south--east direction from the center of M31 and is seen around an extension of the GSS. 
The gravitational pull of M33 might deform the outer density profile of M31 in the past. 
The morphology of the satellite could also change the debris configuration (see \cite{Mik2014b}).
Furthermore, \citet{Mik2014a} performed a large set of parameter study to evaluate possible parameter space upon the accreting orbit of the progenitor satellite galaxy. 
They found that the satellite orbit must remain within a narrow parameter range including the one adopted in this paper. 
In addition, \citet{Far2013mnras} also states that the possible parameter space of the initial orbit of the progenitor and the mass of DMH in M31 is narrow to reproduce the GSS. 
Therefore, we think that the different orbit would not be able to solve the disagreement between our result and the CDM prediction. It is, however, necessary to demonstrate in future studies. 

The CDM cosmology excellently matches observations on large scales ($>$ a few Mpc). 
However, some serious discrepancies between CDM predictions and observations are being discussed on smaller scales ($<$ a few Mpc). 
For instance, there is the missing satellites problem, that is the CDM model predicts larger numbers of satellite galaxies than are observed \citep{Nav1996apj,Moo1999mnras}. 
Another well--known issue is the core--cusp problem, that is the mismatch of the observationally inferred central density structures of DMHs when compared with theoretical predictions \citep{1994Natur.370..629M,1995ApJ...447L..25B, Ogi2014mnras}. 
Despite many efforts have been done to solve these problems both within and beyond a framework of the CDM model, we do not yet reach the final conclusion. 
Moreover, we have pointed out that the CDM model faces another issue of the discrepancy in the outer structure of the DMH, and further studies are needed to solve this problem.

\bigskip
We thank M.J. Irwin for using observational data and an anonymous referee for fruitful suggestions that help to improve the paper. 
This work was partially supported by the JSPS Grant--in--Aid for Scientific Research (S)(2024002), (A)(21244013), and (C)(25400222).


\end{document}